\documentclass[pra,twocolumn,nofootinbib,eqsecnum,floatfix,showpacs]{revtex4}
\usepackage{bm} 
\usepackage{amsmath}

\begin{document}

\title{{\bf The Born Rule Dies}
\thanks{Alberta-Thy-04-09, arXiv:0903.4888 [hep-th]}}

\author{Don N. Page}
\email{don@phys.ualberta.ca}

\affiliation{Theoretical Physics Institute\\
Department of Physics, University of Alberta\\
Room 238 CEB, 11322 -- 89 Avenue\\
Edmonton, Alberta, Canada T6G 2G7}

\date{2009 March 27}

\begin{abstract}

The Born rule may be stated mathematically as the rule that
probabilities in quantum theory are expectation values of a complete
orthogonal set of projection operators.  This rule works for single
laboratory settings in which the observer can distinguish all the
different possible outcomes corresponding to the projection operators. 
However, theories of inflation suggest that the universe may be so large
that any laboratory, no matter how precisely it is defined by its
internal state, may exist in a large number of very distantly separated
copies throughout the vast universe.  In this case, no observer within
the universe can distinguish all possible outcomes for all copies of the
laboratory.  Then normalized probabilities for the local outcomes that
can be locally distinguished cannot be given by the expectation values
of any projection operators.  Thus the Born rule dies and must be
replaced by another rule for observational probabilities in cosmology. 
The freedom of what this new rule is to be is the measure problem in
cosmology.  A particular volume-averaged form is proposed.

\end{abstract}


\maketitle

\section{Introduction}

In quantum theory, all probabilities for a system are supposed to be
encoded in its quantum state.  This may be true, but there is the
question of how to decode the quantum state to give these
probabilities.  In traditional quantum theory, using the statistical
interpretation of the wavefunction given by the Nobel Prize-winning Born
rule \cite{Born}, the probabilities are given by the expectation values
of projection operators.  Once a possible observation is specified
(including the corresponding projection operator), then its probability
is given purely by the quantum state as the expectation value the state
assigns to the projection operator.

This prescription seems to work well in ordinary single laboratory
settings, where the observer can distinguish between a set of different
outcomes.  Then distinct observations are mutually exclusive, so that
different ones cannot both be observed.  If one assigns a projection
operator to each possible distinct observation in a complete exhaustive
set, then these projection operators will be orthonormal, and their
expectation values will be nonnegative and sum to unity, which are
conditions necessary for them to be interpreted as the probabilities of
the different possible observations.

However, in cosmology there is the possibility that the universe is so
large that there are many copies of each laboratory and observer, no
matter how precisely the laboratory and observer are defined.  (Here
what one interprets to be a `laboratory' can be as small or as large as
one would like, perhaps as small as a single particle or as large as the
entire earth or solar system, or maybe even as large as an entire causal
diamond \cite{Bousso}.  Similarly, the observer can be as small or as
large as one might consider an observer to be, perhaps as small as
whatever part of the brain of a tiny animal leads to a single conscious
perception, or as large as a whole community, such as a human scientific
information gathering and utilizing system \cite{HS}, or even as large
as the set of all humans and other conscious beings within the solar
system.  For this discussion, however, I would like to restrict to cases
in which the different parts of the `observer' are in communication with
each other, so that I shall not count as a single `observer' collections
of more than one entirely separate civilization, not in communication
with each other, say very widely separated within the universe so that
they actually or effectively have no causal contact with each other.)

This possibility of many identical copies of each laboratory and
observer raises the problem \cite{HS,typdef,typder,cmwvw,insuff} that
two observations that are seen as distinct for an observer are not
mutually exclusive in a global viewpoint of the entire universe; both
can occur for different copies of the laboratory and observer (though
neither copy may be aware of that).  This would not be a problem for a
putative superobserver who can observe all possible sets of observations
by all observers over the entire universe, but it is a problem for the
assignment of normalized probabilities for the possible observations
that are distinct for each copy of the observer.  The result
\cite{typder,insuff} is that one cannot get such a set of normalized
probabilities as the expectation values of projection operators in the
full quantum state of the universe.

One can use Tegmark's language \cite{Teg,AT,HAT} of a `bird' for a
superobserver that can `observe' (not necessarily in the traditional
quantum sense of making a measurement that disturbs the system) or know
the entire history of the universe (or its entire quantum state), and a
`frog' for a localized observer that is entirely in communication or
causal contact with itself within the universe.  The problem is that
traditional quantum theory with its Born rule applies to birds but not
to frogs, if the universe is so large that there are many exact copies
of the frogs.  In other words, traditional quantum mechanics is for the
birds.

What we need instead is a replacement of Born's rule that applies within
the universe to us frogs that might exist in many copies, identical
except for our location relative to our distant surroundings that we are
not aware of.  (If different frogs are aware that their surroundings are
different, that awareness makes the frogs themselves different and so
not exact copies of each other.  But if the differences in the
surroundings are not reflected in the internal states of the frogs, say
if the surroundings are so far away that they are not in causal contact
with the frogs, then the frogs can be identical and yet be at different
locations as seen by the birds.) 

One can still postulate that there are rules for getting the
probabilities of all possible frog observations from the quantum state,
but then the question arises as to what these rules are.  Below we shall
give examples of several different possibilities for these rules,
showing that they are not uniquely determined.  Thus, they are logically
independent of the question of what the quantum state is.  Therefore,
the quantum state just by itself is insufficient to determine the
probabilities of observations for us frogs within the universe.

The main application of the logical independence of the probability
rules is to the {\it measure problem} in cosmology \cite{BLL94,Vil95a,
LLM,GL,LLM2,Vil95b,WV,LM,Vil98,VVW,Guth00,GV,EKS,SEK,AT,Teg2,
Aguirre,Ellis,Page-in-Carr,GSVW,ELM,BFL,Bou06,CDGKL,BFY06,GT,
Vil07,AGJ1,Win06,AGJ2,BHKP,Guth07,BY,LWa,ADNTV,Linde08a,Linde08b,
LWb,CSS,Haw,GV07,HHH,Win08a,SGSV,BFYb,Win08b,DGLNSV,cmwvw,
Win09c,LVW,GV09,Bou09},
the problem of how to make statistical predictions for observations in a
universe that may be so large that almost all theoretically possible
observations actually occur somewhere.  The logical independence implies
that the solution to the measure problem is not just the quantum state
of the universe but also other independent elements, the rules for
getting the probabilities of the localized (frog) observations from the
quantum state.

\section{The set of all possible observations}

One goal of science is to produce good theories $T_i$ that enable one to
predict the probabilities of results of observations by localized
observers within the universe (`frogs', rather than `birds' that are
hypothetical superobservers for the entire universe).  For brevity, I
shall refer to the results of observations simply as `observations.' 
Here for simplicity I shall assume that there is a countable set of
possible distinct observations $O_j$ out of some exhaustive set of all
such observations. 

If one imagines a continuum for the set of observations (which seems to
be logically possible, though not required), in that case I shall assume
that they are binned into a countable number of exclusive and exhaustive
subsets that each may be considered to form one distinct observation
$O_j$.  Then the goal of a good complete theory $T_i$ is to calculate
the probability $P_j(i) \equiv P(O_j|T_i)$ for each possible observation
$O_j$, given the theory $T_i$.

The set of possible observations might be, for example, all possible
conscious perceptions \cite{SQM,Page-in-Carr}, all possible data sets
for one person, all possible contents for an eprint arXiv, all possible
data sets for a human scientific information gathering and utilizing
system \cite{HS}, or all possible data sets for any community of
observers.  However, as mentioned above, I shall assume that the
observer, whether a single organism or a whole community, is in complete
causal contact with itself and is not a collection of disconnected
organisms or communities.

In order that the sum of the probabilities for all members of the set of
possible observations be unity, I shall also assume that the set of
observations is mutually exclusive, so that any particular observation
is a unique and distinct member of the set.  In particular, each
observation is to be complete and not a subset of another observation,
to avoid double counting and unnormalized probabilities.  The simplest
way I know to impose this is to take observations to be conscious
perceptions, all that an organism is phenomenally aware of at once
\cite{SQM,Page-in-Carr}.  I shall generally have in mind this example of
what an observation is, but my argument applies to any definition of an
observation so long as the possible observations are mutually exclusive.

For example, if one defined observations to be data obtained by or held
by a community of observers rather than by a single organism, one needs
to define what comprises the community so that subsets of the community
are not separately counted.  That is, suppose one has organisms $A$,
$B$, and $C$ with observed data $\alpha$, $\beta$, and $\gamma$
respectively.  One could define communal observers to be any subset of
the set $\{A,B,C\}$ of organisms and observations to be the data of that
subset. One allowed choice of what the observations are would be
$\{\alpha,\beta,\gamma\}$, with these three distinct observations
(analogous to my personal preference for taking the observations to be
conscious perceptions, assuming that each conscious perception can be
attributed to one or another of the organisms but not jointly to any
combination of them).  Another allowed choice would be to take the
entire community of $A$, $B$, and $C$ as one observer, so that the
observation would be the combined data of $\alpha$, $\beta$, and
$\gamma$, making one single observation, the one-element set
$\{\alpha\beta\gamma\}$.  But in this paper I do not allow one to
consider sets of observations such that one observation within the set
is part of another observation within the set, so that the two cannot be
considered mutually exclusive.  That is, I do not allow the set of
observers to be taken to be all possible nonzero subsets of the set of
the three organisms, the set $\{A,B,C,AB,AC,BC,ABC\}$ with observations
$\{\alpha,\beta,\gamma,\alpha\beta,\alpha\gamma,\beta\gamma,
\alpha\beta\gamma\}$, since, for example, the observation $\alpha$ is
part of the observation $\alpha\beta$, so that these two observations
are not mutually exclusive.

Thus for this paper, the set of possible (frog) observations must be
distinct, so that one observation is not simply part of another
observation also considered to be within the set.  This requirement
seems simple to meet if the observations are considered to be conscious
perceptions, since if organism $A$ has conscious perception $\alpha$ and
organism $B$ has conscious perception $\beta$, the combination
$\alpha\beta$ is not a conscious perception but a combination of two. 
(Even for a single organism, its conscious perceptions at two different
times are distinct conscious perceptions, and the combination of the two
is not a conscious perception, an awareness that is perceived at once by
any frog within the universe.  When one is aware of both the past and
present, this is not actually a combination of a past conscious
perception and a present conscious perception, but a single present
conscious perception that has elements within it both of awareness of
the present and of awareness of memories of the past.)  This paper
allows observations to be more general than conscious perceptions (for
those readers sceptical about the fundamental nature of conscious
perceptions), but for the conclusions of this paper, one is not allowed
to consider a set of possible observations such that one element of the
set contains another element rather than being distinct from it.

To give another illustration of the restriction, consider the idea that
observations are finite strings of binary digits within a toy universe
made up of an infinite string of binary digits.  It would be acceptable
to take the set of possible observations to be the set of all binary
strings of a certain fixed length, since all such strings would be
distinct.  However, if one tried to take the set of possible
observations to be the set of all binary strings of all finite lengths,
then the observations would not be distinct, since, for example, the
string $\{0101\}$ would be contained within the string $\{01010\}$ and
so would not be distinct:  if one observation was of $\{01010\}$, then
the observation $\{0101\}$ would also necessarily occur.

This example does not imply that one could not consider the set of
possible observations to include strings of different length, so long as
the length is part of the observation.  For example, one could define
the set of observations to be all finite substrings of the universe's
infinite string that start and end with the string $00$ and have no $00$
inside (between the two $00$'s at the ends).  Then the set of possible
observations would be the strings $\{0000\}$, $\{00000\}$, $\{00100\}$,
$\{000100\}$, $\{001000\}$, $\{001100\}$, etc. (with all other strings
being longer than these ones explicitly listed and with $\{000000\}$ not
being a possible observation since it has a $00$ between the two $00$'s
at the ends).  The idea is that if in this toy model possible
observations are defined to be certain strings of digits or integers,
then no allowed strings should be proper subsets of other allowed
strings.

Note that in this paper I am not assuming that observations are
eigenstates or eigenvalues of Hermitian operators, or that they
correspond to subspaces of a Hilbert space.  Locally they may have that
form, but the fact that the location of the observation is indeterminate
(cannot be known by the frogs making the observation) means that
observations are not globally eigenstates or eigenvalues or subspaces of
a Hilbert space.  Therefore, theorems such as Gleason's \cite{Gleason}
that may be taken to imply the uniqueness of Born's rule need not apply
to the frog observations being considered in this paper.

\section{Probabilities for observations}

Once one has defined a mutually exclusive set of all possible
observations $O_j$, a goal of science is to produce good theories $T_i$
that each give normalized nonnegative probabilities $P_j(i) \equiv
P(O_j|T_i)$ for the observations $O_j$,
\begin{equation}
\sum_j P_j(i) = 1, 
\label{norm}
\end{equation}
for each theory $T_i$.

One might think that once one has the quantum state, there would be a
standard answer to the question of the probabilities for the various
possible observations.  For example \cite{typder,insuff}, one might take
traditional quantum theory (what I there called standard quantum theory)
to use Born's rule and hence give the probability $P_j(i)$ of the
observation as the expectation value, in the quantum state given by the
theory $T_i$, of a projection operator $\mathbf{P}_j$ onto the
observational result $O_j$.  That is, one might take
\begin{equation}
P_j(i) = \langle \mathbf{P}_j \rangle_i, 
\label{Born}
\end{equation}
where $\langle \rangle_i$ denotes the quantum expectation value of
whatever is inside the angular brackets in the quantum state $i$ given
by the theory $T_i$.  This traditional approach (Born's rule) works in
the case of a single laboratory setting where the projection operators
onto different observational results are orthogonal, $\mathbf{P}_j
\mathbf{P}_k = \delta_{jk}\mathbf{P}_j$ (no sum over repeated indices).

However \cite{typder,cmwvw,insuff}, in the case of a sufficiently large
universe, one may have many laboratories that are locally identical
(e.g., without consideration of the surroundings that are not at all
reflected by the quantum state or data within the laboratory, data that
are accessible only to a hypothetical superobserver or bird).  Then,
within different copies of these locally identical laboratories
(different only from the bird perspective), one can have observation
$O_j$ occurring `here' and observation $O_k$ occurring `there' in a
compatible way, so that $\mathbf{P}_j$ and $\mathbf{P}_k$ are not
orthogonal.  Then the traditional quantum probabilities given by Born's
rule, Eq.\ (\ref{Born}), will not be normalized to obey Eq.\
(\ref{norm}).  An explicit proof of this will now be given for a toy
model.

\section{Toy model proof that Born's rule does not work}

To illustrate the problem with Born's rule and prove that we cannot obey
Eq.\ (\ref{norm}) with Born-rule probabilities Eq.\ (\ref{Born}),
let us consider a toy model for a universe at one moment of time in
which each component of the quantum state has $N$ regions that can each
have either no observer, denoted by 0, or one observer (which, as
discussed above, can be taken to be an entire communicating community,
so long as it is defined so that proper subsets of the community are not
also counted as additional observers) with observation $O_j$, denoted by
$j$.  (Different values of $N$ model different sizes of universes
produced by differing amounts of inflation in the cosmological measure
problem.)  Write the quantum state as a superposition, with complex
coefficients $a_N$, of component states $|\psi_N\rangle$ that each have
different numbers $N$ of regions:
\begin{equation}
|\psi\rangle = \sum_{N=1}^{\infty} a_N |\psi_N\rangle, 
\label{state}
\end{equation}
where $\langle\psi_M|\psi_N\rangle = \delta_{MN}$.  Let each of these
component states $|\psi_N\rangle$ of fixed size (number of regions being
$N$) be called a size eigenstate, or an $N$-state if we want to denote
the size or number of regions $N$.

Furthermore, write each component state with a definite number $N$ of
regions as a superposition of orthonormal states in the tensor product
of $N$ regions that can each be labeled by either having no observation,
$0$, or by having the observation $j$, in the region $L$, $1 \leq L \leq
N$.  That is, if we let $\mu_L$ be either $0$ if the region $L$ has no
observer or else $j$ if the region $L$ has the observation $j$, then the
state for a definite $N$ can be written as
\begin{equation}
|\psi_N\rangle = \sum_{\mu_1,\mu_2,\ldots,\mu_N}
 b_{\mu_1\mu_2\ldots\mu_N} |\mu_1\mu_2\ldots\mu_N\rangle, 
\label{N-state}
\end{equation}
where the component state $|\mu_1\mu_2\ldots\mu_N\rangle$ has $\mu_1$
(either no observation, 0, or the observation $O_j$ that is denoted by
$j$) in the first region, $\mu_2$ in the second region, and so on with
all $\mu_L$ for $0\leq L\leq N$ up through $\mu_N$ in the $N$th region. 
Let each of these component states be called a bird's-eye eigenstate of
the observations, or a bird eigenstate.  (Each frog can only know of its
particular $O_j$ and not what the bird eigenstate is or even what its
own location $L$ is within the bird eigenstate with its $N$ regions.) 
The full quantum state
$|\psi\rangle$ is then a superposition of size eigenstates
$|\psi_N\rangle$ given by Eq.\ (\ref{state}), with each size eigenstate
being itself a superposition of bird eigenstates
$|\mu_1\mu_2\ldots\mu_L\rangle$ given by Eq.\ (\ref{N-state}).

It is also convenient to define an $NN_j$ eigenstate as the sum, for
fixed $N$, given by Eq.\ (\ref{N-state}) restricted to the terms that
have a fixed number $N_j$ of occurrences of the observation $j$ within
the string of regions.  One can further define an $NN_jN_k$ eigenstate
as the sum, for fixed $N$, given by Eq.\ (\ref{N-state}) restricted not
only to the terms that have a fixed number $N_j$ of occurrences of the
observation $j$ but also to terms that have the fixed number $N_k$ of
the observation $k$, $j\neq k$.  One could then include just these sums
for each $N$ in the sum over $N$ given by Eq.\ (\ref{state}) to get the
corresponding $N_j$ and $N_jN_k$ eigenstates that are eigenstates of the
occurrence of precisely $N_j$ observations $j$ and of the mutual
occurrence of precisely $N_j$ observations $j$ and precisely $N_k$
observations $k$.  Of course, distinguishing all of these eigenstates is
only possible for hypothetical superobservers or birds who can see the
entire universe; it is not possible for the frogs that can see only what
is in their individual regions.

Now let us assume these very minimal principles:

{\it No Extra Vision Principle\/} (NEVP):

The probabilities of observations that have zero amplitudes to occur
anywhere in the quantum state are zero; one cannot see what is not there
in the quantum state.

{\it Probability Symmetry Principle\/} (PSP):

If the quantum state is an eigenstate of equal number of observations of
two different observations, then the probabilities of these two
observations are equal.

The No Extra Vision Principle is the assumption that frogs can't see
what birds don't see, that if for some $j$ the bird quantum probability
$p_{NLj}$ is zero for all $N$ and $L$ in theory $T_i$, then the frog
probability $P_j(i) \equiv P(O_j|T_i)$ is also zero.  In other words, I
assume that if there is zero quantum probability for the bird to see the
observation occurring anywhere, then the probability is also zero for
the frogs to make that observation.

If one did not make some assumption rather like that, it is hard to see
how the quantum state could determine the frog probabilities in a
reasonable way.  I suppose one might take a mystical attitude and
postulate that frogs can see visions of what has no basis in the
physical quantum state, but for this paper I shall assume that anything
that can be observed within the universe has a corresponding amplitude
in the quantum state.

(Note that this assumption does not rule out visions or even revelations
from God that do not directly correspond to physical reality external to
the observer, since they can be caused by the part of the quantum state
of the observer himself/herself/itself.  In principle that part of the
quantum state could even be directly caused by God Himself without
having to go through the usual mechanism of coming in from external
stimuli, though of course external stimuli can also be directly caused
by God and can be so even in the case in which they are correctly
described by theories of physics; the two descriptions are not logically
contradictory.  Nothing in this paper assumes or denies the possibility
of such visions and/or revelations, but I am assuming that whatever is
observed does have a corresponding nonzero amplitude in the quantum
state to mediate it.  This might be taken to be an analogue in physics
of the Biblical claim that the Word became flesh.)

The Probability Symmetry Principle is the assumption that if the full
quantum state $|\psi\rangle$ is an $N_jN_k$ eigenstate with $N_j=N_k$ in
theory $T_i$, then $P_j(i)=P_k(i)$ for $j\neq k$ (though this equality
is also trivially true for $j=k$).  The idea of this principle is that
if the quantum state has equal definite numbers of two different
observations, there seems to be no good reason to assign one of them
greater probability than the other.  The only thing that distinguishes
them (other than what the observations are intrinsically) is their
location, and it is hard to see why that should favor one over another. 
Furthermore, if one imagined some sort of diffeomorphism invariance that
allowed one to translate any one location to any other, there would be
no absolute distinction to the locations, so why should one assign
different probabilities to observations at the different locations?  (Of
course, one could say that from the bird's eye view, though not to the
frogs, the relative locations can differ, so it might be that one of the
observations of $j$ is next to another observation of $j$, whereas there
is no observation of $k$ next to another observation of $k$, but it is
hard to see why this nonlocal relative information that only the birds
have should affect the probabilities of the local frog observations.)

To show that Born's rule, Eq.\ (\ref{Born}), necessarily violates the
normalization condition, Eq.\ (\ref{norm}), when the No Extra Vision
Principle and the Probability Symmetry Principle are assumed, consider
the size eigenstates with $N=2$ that have the form
\begin{equation}
|\psi\rangle =  b_{11}|11\rangle + b_{12}|12\rangle
              + b_{21}|21\rangle + b_{22}|22\rangle, 
\label{2-state}
\end{equation}
so that in each component, precisely two observations are made that are
either $j=1$ or $j=2$, with the state normalized, $|b_{11}|^2 +
|b_{12}|^2 + |b_{21}|^2 + |b_{22}|^2 = 1$ .

Now if $b_{11} = 1$, so $|\psi\rangle = |11\rangle$ (an $N_jN_k$
eigenstate with $N_1=2$ and $N_2=0$), then only the observation 1 occurs
as seen by the birds, so $P_1 = 1$ and $P_2=0$ by the No Extra Vision
Principle and by the normalization requirement.  Similarly, if $b_{22} =
1$, so $|\psi\rangle = |22\rangle$ (an $N_jN_k$ eigenstate with $N_1=0$
and $N_2=2$), then only the observation 2 occurs, so then $P_1 = 0$ and
$P_2=1$.  On the other hand, a state orthogonal to both of these
$N_jN_k$ eigenstates, $|\psi\rangle = b_{12}|12\rangle +
b_{21}|21\rangle$ with $b_{11} = b_{22} = 0$ is an $N_jN_k$ eigenstate
with $N_1=N_2=1$ and so by the Probability Symmetry Principle has
$P_1=P_2$, both of which must be $1/2$ to be normalized, obeying Eq.\
(\ref{norm}).

The proof that Born's rule fails to give this proceeds by showing that
there is no pair of orthogonal projection operators $\mathbf{P}_1$ and
$\mathbf{P}_2$ in this 4-state system whose expectation values reproduce
these probabilities by Born's rule Eq.\ (\ref{Born}) for all such
quantum states.

For Born's rule to give, for $|\psi\rangle = |11\rangle$, $P_1 = \langle
\mathbf{P}_1 \rangle \equiv \langle\psi|\mathbf{P}_1|\psi\rangle =
\langle 11|\mathbf{P}_1|11\rangle = 1$, the projection operator
$\mathbf{P}_1$ must have the form $|11\rangle\langle 11| +
\mathbf{P}'_1$ where $\langle 11|\mathbf{P}'_1|11\rangle = 0$. 
Similarly, for Born's rule to give, for $|\psi\rangle = |22\rangle$,
$P_2 = \langle \mathbf{P}_2 \rangle \equiv
\langle\psi|\mathbf{P}_2|\psi\rangle = \langle 22|\mathbf{P}_2|22\rangle
= 1$, the projection operator $\mathbf{P}_2$ must have the form
$|22\rangle\langle 22| + \mathbf{P}'_2$ where $\langle
22|\mathbf{P}'_2|22\rangle = 0$.

Furthermore, for Born's rule to give, for $|\psi\rangle = |11\rangle$,
$P_2 = \langle 11|\mathbf{P}_2|11\rangle = 0$, we need $\langle
11|\mathbf{P}'_2|11\rangle = 0$.  Similarly, for Born's rule to give,
for $|\psi\rangle = |22\rangle$, $P_1 = \langle 22| \mathbf{P}_1
|22\rangle = 0$, we need $\langle 22| \mathbf{P}'_1 |22\rangle = 0$. 
Since $\mathbf{P}_1$ and $\mathbf{P}_2$ are to be orthogonal projection
operators, $\mathbf{P}_1\mathbf{P}_1 = \mathbf{P}_1$,
$\mathbf{P}_2\mathbf{P}_2 = \mathbf{P}_2$, and $\mathbf{P}_1\mathbf{P}_2
= \mathbf{P}_2\mathbf{P}_1 = 0$, one can easily show that
$\mathbf{P}'_1$ and $\mathbf{P}'_2$ themselves must be orthogonal
projection operators within the subspace of states of the form
$b_{12}|12\rangle + b_{21}|21\rangle$ that are orthogonal to
$|11\rangle$ and $|22\rangle$.

That is, to give the right probabilities for the states $|\psi\rangle =
|11\rangle$ and $|\psi\rangle = |22\rangle$ by Born's rule, and to give
nonzero probabilities when $N_1=N_2=1$, the projection operators must
have the form $\mathbf{P}_1 = |11\rangle\langle 11| +
|\psi_{12}\rangle\langle\psi_{12}|$ and $\mathbf{P}_2 =
|22\rangle\langle 22| + |\psi'_{12}\rangle\langle\psi'_{12}|$ where
$|\psi_{12}\rangle = \cos{\theta}|12\rangle +
\sin{\theta}e^{i\phi}|21\rangle$ and $|\psi'_{12}\rangle =
-\sin{\theta}e^{-i\phi}|12\rangle + \cos{\theta}|21\rangle$ are two
orthogonal states within the subspace orthogonal to $|11\rangle$ and
$|22\rangle$, with the two arbitrary real parameters $\theta$ and
$\phi$.

So far we have not run into the problem, but now we do:  For a generic
$N_jN_k$ eigenstate with $N_1=N_2=1$, $|\psi\rangle = b_{12}|12\rangle +
b_{21}|21\rangle$, Born's rule does not give $P_1 = \langle \mathbf{P}_1
\rangle = 1/2$ or $P_2 = \langle \mathbf{P}_2 \rangle = 1/2$.  In
particular, if the quantum state has $|\psi\rangle = |\psi_{12}\rangle$,
then $\langle \mathbf{P}_1 \rangle = 1$ and $\langle \mathbf{P}_2
\rangle = 0$ instead of $P_1 = P_2 = 1/2$ for this $N_jN_k$ eigenstate
as required by the Probability Symmetry Principle.

One might object that if one were allowed to choose the projection
operators $\mathbf{P}_1$ and $\mathbf{P}_2$ after the state is known,
one can avoid the problem altogether.  For example, if one knows that
the quantum state is the particular $N_jN_k$ eigenstate with $N_1=N_2=1$
that is $|\psi\rangle = b_{12}|12\rangle + b_{21}|21\rangle$, then with
any orthogonal state $|\psi'\rangle = b^*_{21}|12\rangle -
b^*_{12}e^{i\varphi}|21\rangle$ one can choose $|\psi_{12}\rangle =
(|\psi\rangle + |\psi'\rangle)/\sqrt{2}$ and $|\psi'_{12}\rangle =
(|\psi\rangle - |\psi'\rangle)/\sqrt{2}$ to get $\mathbf{P}_1 =
|11\rangle\langle 11| + |\psi_{12}\rangle\langle\psi_{12}|$ and
$\mathbf{P}_2 = |22\rangle\langle 22| +
|\psi'_{12}\rangle\langle\psi'_{12}|$ that would then give $\langle
\mathbf{P}_1 \rangle = \langle \mathbf{P}_2 \rangle = 1/2$, in accord
with the Probability Symmetry Principle.  However, it is surely not
correct to have to know the quantum state in order to choose the
projection operators whose expectation values in that state are the
probabilities.  If one were allowed to do that, one could always choose
the orthonormal projection operators to give any normalized set of
nonnegative probabilities one wanted, no matter what the quantum state
is.  In order for the probabilities to depend on the quantum state in
some reasonable way, the rule for extracting the probabilities from the
state should not be allowed to depend on the quantum state in such an
{\it ad hoc} manner.

One might also object to the Probability Symmetry Principle, but it
seems highly unnatural to give that up.  One could instead just pick a
preferred location and then normalize the relative probabilities for the
various observations to occur there.  Even if one had diffeomorphism
invariance so that intrinsically there is no preferred location, one
might just choose the relative location that, say, has the smallest rms
distance to all other observations, as seen by the bird.  But such {\it
ad hoc} choices to favor one location over another certainly seem
unlikely to be the way that nature really works, though one might want
to consider it further before rejecting it completely.

Thus we have seen that if we require that the projection operators be
chosen independently of the quantum state, and if the observational
probabilities are to obey the Probability Symmetry Principle (equal
probabilities for different observations that definitely occur an equal
number of times), such probabilities cannot be given by Born's rule as
the expectation value of the projection operators, Eq.\ (\ref{Born}), if
we allow states with more than one observation actually occurring that
are not just quantum alternatives.

Thus one needs a formula different from Born's rule for normalized frog
probabilities of a mutually exclusive and exhaustive set of possible
observations, when distinct observations within the complete set cannot
be described by orthogonal projection operators.  Here, by an
`exhaustive' set of possible observations, I mean all that can be
locally distinguished by the frogs, without considering the distinctions
that the birds may see by being able to identify the different locations
(different distant surroundings) of the copies of the laboratories and
observers.

\section{Comparison with classical theory}

One might note that conceptually this problem is not peculiar to quantum
theory.  If one has a classical theory, one might say that the analogue
of the expectation values of projection operators is the set of 0's and
1's for whether something does not occur or does occur in the classical
behavior.  One might then say that the analogue of Born's rule in
classical theory is that the probability is 0 for something that does
not occur and 1 for something that does occur.  This is fine for the
view of a bird that can see the entire classical behavior and can see
whether or not something occurs.  However, for the view of a frog within
the classical universe that cannot see the whole thing, one wants
normalized probabilities for the frog observations.  If more than one
observation can occur within the classical behavior, then they cannot
have normalized probabilities that are each 1.  One might take the
number of times a particular observation occurs (or the time interval
during which it occurs, if it occurs continuously over a finite range of
time) and divide by the total for all possible observations to get a
normalized probability for the particular observation, but if there are
more than one actual observations, this normalized probability for the
observation is not 1 and so is not given by the classical analogue of
Born's rule as the expectation value of a projection operator.

Perhaps the problem is so obvious in classical theory that few would
propose that normalized probabilities for a set of possible observations
in classical theory would all be 0's or 1's if they considered cases in
which more than one observation can actually occur.  But in quantum
theory, the expectation values of projection operators can be anywhere
between 0 and 1 inclusive, so perhaps it was not so obvious that there
also one cannot use these expectation values as normalized probabilities
for observations if more than one can actually occur.

Another reason why this problem was not widely recognized is that what
we can see of the universe is so small, in some sense, that most
macroscopic systems within it are complex enough to be unique within the
limited part of the universe we can observe.  Atoms and small molecules
of course are not complex enough to be unique within what we can see; we
believe there are billions of identical atoms and molecules within each
of us.  However, even objects as small as snowflakes have far, far more
possible configurations (presumably very roughly the exponential of some
not-too-small fractional power, perhaps 2/3, of the number of molecules,
say $10^{20}$, in a snowflake, I would guess at least $\exp{10^{10}}$
possible molecular configurations if the fractional power is not smaller
than one half) than the number of particles in the observable part of
the universe (which is much, much less than $\exp{10^{2.5}}$, which is
roughly the exponential of the one-eighth power of the number of
molecules in a snowflake; surely the fractional power is not so small as
one eighth).  Therefore, it is plausible that no two snowflakes in our
part of the universe are identical.

(There may be some snowflake configurations, such as some precisely
regular crystals, that are so much more probable than the average
irregular configurations that it is conceivable that more than one of
them might occur within the observable universe.  I simply do not know
the probabilities for such perfect crystals with a definite symmetric
arrangement of water molecules, which by minimizing the energy would
maximize the probability for a microscopic configuration in a thermal
state.  However, it is surely the case that most snowflakes are
sufficiently irregular that it is highly improbable that any others of
those have the same microscopic arrangements of molecules, say using the
quantum criterion that the trace of the product of the density matrices
for the relative locations of the hydrogen and oxygen nuclei within the
two snowflakes is greater than, say, one half, to avoid the objections
of those who might say no two snowflakes at different locations would
have {\it exactly} the same quantum state in the sense of having {\it
exactly} the same density matrices, with {\it exactly} the same
probabilities of all the excited energy eigenstates.)

With much of what we can see being very likely unique within our
observable universe, when one ignores the possibility that the universe
extends far beyond what we can see, one can also ignore the possibility
that our observation, at least if it is sufficiently detailed, occurs
more than once.  However, when we consider the possibility that the
universe is exponentially larger than what we can see of it (either from
a model of an infinite universe, such as a simply connected $k=-1$
Friedmann-Robertson-Walker model, or from a finite model that has been
expanded to enormously large sizes by inflation), we must face the
possibility that a local observation by a frog, no matter how detailed,
may not be unique from the bird perspective.  Then from the bird
perspective, the different observations are not mutually exclusive, so
the sum of their probabilities within the bird perspective can exceed
unity.  However, from the frog perspective, its observation is a unique
realization out of the set of possible frog observations, so we frogs
want normalizable probabilities for our possible frog observations. 
This discrepancy between the exclusivity of the observations from the
frog perspective and their mutual compatibility from the bird
perspective is what prevents the frog probabilities from all being 0 or
1 in classical theory and from being expectation values of orthonormal
projection operators in quantum theory.  It may be that the Born rule
works for the birds, but it does not work for us frogs.

\section{Examples of different observational probabilities for the same
quantum state}

Let us demonstrate the logical freedom in the rules for the
observational probabilities $P_j(i) \equiv P(O_j|T_i)$ by exhibiting
various examples of what they might be. For simplicity, let us restrict
attention to theories $T_i$ that all give the same pure quantum state
$|\psi\rangle$, which in the toy model above with different regions at
one moment of time can be written as the superposition given by Eqs.\
(\ref{state}) and (\ref{N-state}), i.e., as
\begin{eqnarray}
|\psi\rangle &=& \sum_{N=0}^{\infty} a_N |\psi_N\rangle \nonumber \\ 
&=& \sum_{N=0}^{\infty} a_N\!\! \sum_{\mu_1,\mu_2,\ldots,\mu_N}\!\!
 b_{\mu_1\mu_2\ldots\mu_N} |\mu_1\mu_2\ldots\mu_N\rangle
\label{full-state}
\end{eqnarray}
in terms of the bird eigenstates $|\mu_1\mu_2\ldots\mu_N\rangle$ in
which the first of the $N$ regions has $\mu_1 = 0$ if no observation
occurs there or $\mu_1 = j$ if the observation $O_j$ occurs there, and
similarly for all other regions $L$ up to $N$ for each $N$.

As an example of such a state that I shall use to compare the results of
the various replacements of Born's rule is the following superposition
of two bird eigenstates of different sizes:
\begin{equation}
|\psi\rangle = \cos{\theta} |m_0;m_1;m_2\rangle 
                +\sin{\theta} |n_0;n_1;n_2\rangle, 
\label{example-state}
\end{equation}
where $|m_0;m_1;m_2\rangle$ means the bird eigenstate with $N_m = m_0 +
m_1 + m_2$ regions such that $\mu_L = 0$ (no observation) for the first
$m_0$ regions, $0 \leq L \leq m_0$, $\mu_L = 1$ (the observation $O_1$)
in the next $m_1$ regions, $m_0 + 1 \leq L \leq m_0 + m_1$, and $\mu_L =
2$ (the observation $O_2$) in the last $m_2$ regions, $m_0 + m_1 + 1
\leq L \leq N_m = m_0 + m_1 + m_2$, and where, similarly,
$|n_0;n_1;n_2\rangle$ means the bird eigenstate with $N_n = n_0 + n_1 +
n_2$ regions such that $\mu_L = 0$ for the first $n_0$ regions, $\mu_L =
1$ in the next $n_1$ regions, and $\mu_L = 2$ in the last $n_2$
regions.  To illustrate some points I wish to make below, unless stated
otherwise I shall assume the generic case in which none of the integers
$m_0,m_1,m_2,n_0,n_1,$ and $n_2$ are zero or are equal to each other.

Let us suppose that $\mathbf{P}_\mu^L$ is a complete set of orthogonal
projection operators for the region $L$, either for there to be no
observation ($\mu=0$) in that region or else for the observation $O_j$
to occur ($\mu=j$) there, so $\mathbf{P}_\mu^L\mathbf{P}_\nu^L =
\delta_{\mu\nu}\mathbf{P}_\mu^L$ (no sum over $\mu$) and $\sum_\mu
\mathbf{P}_\mu^L = I$, the identity operator.  (In a more realistic
model, I would not assume that any of these projection operators is rank
one, having only pure state eigenstates, so they would not pick out a
unique basis, but in my toy model I am assuming that the state in each
region is given uniquely by the value of $\mu$ there for the bird
eigenstates.)  For simplicity, I am also assuming that all of the
different regions are spacelike separated (e.g., are at the same time),
so that the Hilbert space for each number of regions $N$ is the tensor
product of the Hilbert spaces for each region.  Thus I am assuming that
each $\mathbf{P}_\mu^L$ (which acts on the full quantum state in the
tensor product) acts nontrivially only on its region $L$ and acts
trivially (as the identity) on all the other regions.  In particular,
this means that not only do all of the $\mathbf{P}_\mu^L$ with the same
$L$ commute with each other by their relation as a complete set of
orthonormal projection operators, but also all of the $\mathbf{P}_\mu^L$
for different $L$ commute with each other as well.

If the quantum state were the size eigenstate $|\psi_N\rangle$, then the
bird quantum probability that the observation $O_j$ occurs in the region
$L$ (in the view of the bird that can tell what the region $L$ is) would
be
\begin{equation}
p_{NLj} = \langle\psi_N|\mathbf{P}_j^L|\psi_N\rangle. 
\label{Lprob}
\end{equation}
However, in reality, even just in the component state $|\psi_N\rangle$
for $N>1$, there are other regions where the observation could occur, so
the frog-view probability $P_j(i)$ for the observation $O_j$ in the
theory $T_i$ can be some $i$-dependent function of all the $p_{NLj}$'s. 
The freedom of this function is part of the independence of the
observational probabilities from the quantum state itself.

The frog probabilities logically need not even be functions of the bird
probabilities.  However, at least in my toy model of a universe with
separate regions at one moment of time, so that all the projection
operators defined above (which the birds can observe) commute, it seems
plausible that they would be.  (It is a further challenge to describe
what happens when frog observations occur at different times, so that
the corresponding projection operators in the bird view do not commute,
but I shall not address this issue here.)

In particular, I shall assume the No Extra Vision Principle, that if for
all $N$ and $L$, the bird quantum probability $p_{NLj}$ in theory $T_i$
is zero for some $j$, then the frog probability $P_j(i) \equiv
P(O_j|T_i)$ is also zero for that $j$.  I shall also continue to assume
the Probability Symmetry Principle.  For my simplified toy example state
given by Eq.\ (\ref{example-state}), the PSP would imply that if $\theta
= 0$, so the state were a single bird eigenstate, and if this state had
$m_1 = m_2$ (equal numbers of regions with the observations $O_1$ and
$O_2$ that are each assumed to occur once and only once within each
respective region), then the frog probabilities for these two different
observations would be equal, $P_1 = P_2 = 1/2$, with the value of 1/2
determined by the fact that in this simplified bird eigenstate there is
zero bird probability for any other observation (and hence, by the NEVP
assumption, zero frog probability $P_j$ also), so $P_1$ and $P_2$ must
sum to unity to obey the normalization condition Eq.\ (\ref{norm}). 

Although I am arguing that Born's rule is not sufficient for determining
the frog probabilities from the quantum state, it is surely the case
that whatever the rule is, the resulting probabilities will depend in
some way on the quantum state as well.  The PSP and NEVP assumptions
above seem to be rather natural minimal requirements for this
dependence.  Now I want to give several examples, $T_1$--$T_9$,
illustrating the freedom of the rule within these minimal restrictions,
and with a subset of these examples, $T_3$--$T_5$, satisfying one other
natural restriction.

A useful procedure for getting normalized frog probabilities $P_j(i)$
for theory $T_i$ is first to define some method for getting unnormalized
nonnegative relative frog probabilities $p_j(i)$ and then simply
normalize them by $P_j(i) = p_j(i)/\sum_k p_k(i)$.  So in the following
examples, I shall give examples of rules for calculating relative frog
probabilities $p_j(i)$ from the bird probabilities $p_{NLj}$ that
require the bird's knowledge of the size $N$ and region label $L$ that
the frogs do not have.  The different indices $i$ will denote different
theories, in this case different rules for calculating the
probabilities, since for simplicity we are assuming that all the
theories have the same quantum state $|\psi\rangle$ and hence the same
set of bird probabilities $p_{NLj}$.

Next, let us turn to different possible examples.

(1) For theory $T_1$, use the No Extra Vision assumption to take the
relative frog probability to be zero, $p_j(1)=0$, if all the $p_{NLj}$
are zero for all $N$ and $L$ for that $j$, so that the birds have zero
quantum probability to see any frogs making the observation $O_j$, but
set $p_j(1)=1$ for each frog observation $O_j$ that has at least one
positive $p_{NLj}$ for that $j$.  One might interpret this $p_j(1)$ to
be the `existence probability' of the observation $O_j$ in the Everett
many worlds interpretation of the quantum state.  That is, theory $T_1$
is essentially taking the Everett many worlds interpretation to imply
that if there is any nonzero amplitude for the observation to occur, it
definitely exists somewhere in the many worlds (and hence has existence
probability unity).  If the total number of observations that have
nonzero amplitudes (at least one positive bird probability $p_{NLj}$ for
each such $j$) is $M_1$, then the normalized frog probabilities in this
theory $T_1$ are $P_j(1)=p_j(1)/M_1$.  This would be the theory that
every observation that actually does exist, as seen by the birds who
look at the entire many-worlds quantum state, is equally probable, and
that the observations that have zero bird quantum probabilities in all
components of the many-worlds quantum state do not exist for the frogs
either and hence have zero frog probabilities $P_j(1)$.

The problem with this theory $T_1$ is that for almost all quantum
states, almost all observations will exist in the Everett sense, making
their number $M_1$ nearly as large as the number of all possible
observations, which I am assuming is enormous.  Then the normalized frog
probabilities $P_j(1)=p_j(1)/M_1$ will all be very tiny, giving
extremely low likelihood to the theory $T_1$.  Thus, in a Bayesian
analysis, unless one assigned this theory a prior probability very near
unity in comparison with the priors one assigned to theories having much
higher likelihoods, it seems that surely this theory would have
extremely low posterior probability.  Therefore, I strongly suspect that
other theories can be constructed that would be much more probably true,
given our ordered observations that do not appear as if they have been
randomly chosen with equal probabilities from a set of nearly all
possible observations.

(2) For theory $T_2$, one might try to hang onto Born's rule as closely
as possible by constructing the global projection operator
\begin{equation}
\mathbf{P}_j = \mathbf{I} - \prod_L (\mathbf{I} - \mathbf{P}_j^L) 
\label{existence}
\end{equation}
and using it in Eq.\ (\ref{Born}) to get, not the normalized frog
probabilities $P_j(2)$ (since these expectation values will not be
normalized), but rather to get the relative frog probabilities $p_j(2) =
\langle \mathbf{P}_j \rangle = \langle\psi|\mathbf{P}_j|\psi\rangle$. 
This would not be the full many-worlds existence probability 0 or 1
described for theory $T_1$ (which would be 0 if $p_j(2)=0$ and 1 if
$p_j(2) > 0$), but it might be regarded as the quantum probability for a
superobserver (a bird) to find that at least one instance of the
observation $O_j$ occurs if one imagines the bird making a quantum
observation of all the frog observers in the quantum state
$|\psi\rangle$.

Indeed, this $p_j(2)$ is essentially in quantum language \cite{typder}
what Hartle and Srednicki \cite{HS} propose for the probability of an
observation (without making the distinction between bird probabilities
and frog probabilities), the quantum probability that the observation
occurs at least somewhere.  This is fine for bird probabilities for the
existence of observations that for them are not mutually exclusive. 
However, because the different $\mathbf{P}_j$'s defined this way are not
orthogonal, the resulting quantum probabilities $p_j(2)$ given by Born's
rule will not be normalized to obey Eq.\ (\ref{norm}).  This lack of
normalization is a consequence of the fact that even though it is
assumed that two different observations $O_j$ and $O_k$ (with $j \neq
k$) cannot both occur within the same region $L$, one can have $O_j$
occurring within one region and $O_k$ occurring within another region. 
Therefore, the existence of the observation $O_j$ at least somewhere is
not incompatible with the existence of the distinct observation $O_k$
somewhere else, so the sum of the bird quantum existence probabilities
$p_j(2)$ is not constrained to be unity.  Thus they cannot be used
directly as the normalized frog probabilities.

However, it would be perfectly legal to interpret the unnormalized
$p_j(2)$'s as relative probabilities for the frogs, and from them
construct the corresponding normalized probabilities $P_j(2) =
p_j(2)/\sum_k p_k(2) = p_j(2)/M_2$ for the frogs, with $M_2 = \sum_k
p_k(2)$.  This is what I am defining theory $T_2$ to do.

This theory $T_2$ seems likely to have a higher likelihood than $T_1$,
since presumably there will be many of the $M_1$ observations that have
$p_j(2)>0$ and hence $p_j(1)=1$ but yet have $p_j(2)\ll 1$, so that $M_2
= \sum_k p_k(2)$ is significantly smaller than $M_1$, at least in
absolute terms (that is, I would assume that generically $M_1 - M_2 \gg
1$ if the number of possible observations is very large), though it
would depend on the quantum state and the set of projection operators
$\mathbf{P}_j^L$ whether or not $M_2$ is relatively much smaller than
$M_1$, $M_2/M_1 \ll 1$.  If indeed $M_2/M_1 \ll 1$, and if our
observations are an example of an $O_j$ with $p_j(2)$ near unity, then
$P_j(2) \gg P_j(1)$, so our observations would assign a much higher
likelihood to theory $T_2$ than to $T_1$.  However, I still suspect that
the universe may be so large that a huge number of observations may have
a bird quantum probability near unity to occur at least somewhere, so
that not only $M_1$ but also $M_2$ would be so large that the resulting
likelihood $P_j(2) = p_j(2)/M_2$ for theory $T_2$ would also be
enormously lower than that for other theories of similar elegance (and
hence presumably to be assigned similar prior probabilities) that one
may be able to construct.

For the next sequence of rules from extracting frog observational
probabilities from a given quantum state, I shall assume the 

{\it Probability Fraction Principle\/} (PFP):

If the quantum state has a definite fraction for the ratio of each
possible observation to the total number of all possible observations,
then the probability of each observation is that fraction.

In a restriction of our toy model in which the quantum state
$|\psi\rangle$ is a single bird eigenstates
$|\mu_1\mu_2\ldots\mu_N\rangle$ with either zero or one definite
observation within each region, let
\begin{eqnarray}
N_j &=& \sum_{L=1}^N p_{NLj} \nonumber \\
    &=& \sum_{L=1}^N\langle\mu_1\mu_2\ldots\mu_N|
               \mathbf{P}_j^L|\mu_1\mu_2\ldots\mu_N\rangle
\label{N_j}
\end{eqnarray}
be the number of observations $O_j$ (the number of regions containing
this particular observation, each region assumed to contain at most one
observation).  Further, let
\begin{eqnarray}
N_O &=& \sum_j N_j = \sum_{L=1}^N \sum_j p_{NLj} \nonumber \\
 &=& \sum_{L=1}^N \sum_j \langle\mu_1\mu_2\ldots\mu_N|
      \mathbf{P}_j^L|\mu_1\mu_2\ldots\mu_N\rangle \nonumber \\
 &=& \sum_{L=1}^N\langle\mu_1\mu_2\ldots\mu_N|
               \mathbf{P}_O^L|\mu_1\mu_2\ldots\mu_N\rangle
\label{N_O}
\end{eqnarray}
be the total number of all observations (the total number of regions $N$
minus the number of regions with $\mu_L = 0$ or no observation), where
$\mathbf{P}_O^L = \sum_j \mathbf{P}_j^L$ is the total projection
operator onto having any observation at all (regardless of which
particular $O_j$ it is) in the region $L$.  Since each bird eigenstate
has a definite value of each $N_j$ and of $N_O$, it has the definite
fraction $f_j = N_j/N_O$ for the observation $O_j$.  If the full quantum
state were only this single bird eigenstate, the Probability Fraction
Principle would imply that the probability of the observation $O_j$ is
that fraction, $P_j = f_j$.

The Probability Fraction Principle also implies that if the full quantum
state is a superposition of different bird eigenstates with the same
fractions $f_j = N_j/N_O$ for each (though the $N_j$ and $N_O$ need not
be the same for all component bird eigenstates, only their ratio), then
$P_j = f_j$.  That is, the quantum state need not be an eigenstate of
all the numbers of observations $N_j$, but only of their ratios.  For
example, the quantum state of Eq.\ (\ref{example-state}) for general
$\theta$ (a superposition of two bird eigenstates,
$|m_0;m_1;m_2\rangle$ with $N_1 = m_1$ and $N_2 = m_2$ and 
$|n_0;n_1;n_2\rangle$ with $N_1 = n_1$ and $N_2 = n_2$, with different
values of $N_1$ and of $N_2$) is not an eigenstate of $N_1$ and of $N_2$
but would be an eigenstate of the fractions $f_1 = N_1/(N_1+N_2)$ and of
$f_2 = N_2/(N_1+N_2)$ if $m_1/n_1 = m_2/n_2$.  For such an eigenstate of
the fractions, the Probability Fraction Principle implies that $P_1 =
f_1$ and $P_2 = f_2$. 

Note that the Probability Fraction Principle implies the Probability
Symmetry Principle but not conversely, so that the Probability Fraction
Principle is a stronger principle.  For example, theories $T_1$ and
$T_2$ satisfy the PSP but not the PFP, as one can see for the
superposition above that is a fraction eigenstate with $f_1 \neq f_2$: 
both $T_1$ and $T_2$ give $P_1 = 1/2 \neq f_1$ and $P_2 = 1/2 \neq f_2$.
This fact might be taken as another possible reason for rejecting
theories $T_1$ and $T_2$ (or for assigning them low prior probabilities,
though I argued above that I suspect they would end up with low
posterior probabilities anyway just from their low likelihoods if indeed
there are a huge number of possible observations that $T_1$ and $T_2$
would assign nearly equal, and hence very low, probabilities $P_j$, with
the particular $P_j$ that the theory assigns to our observation to be
used as the likelihood of the theory in a Bayesian analysis).

After imposing the Probability Fraction Principle, the main remaining
freedom in the rule for assigning observation probabilities from the
quantum state is how to weight different components of the quantum state
with different fractions $f_j$ that have different numbers of regions
$N$ and/or different numbers of observations $N_O$ (and hence also with
possibly different numbers $N-N_O$ of regions with no observations). 
For example, different weightings below can correspond to the difference
between using volume weighting or not in inflationary cosmology
\cite{cmwvw}.

(3) For theory $T_3$, corresponding to volume weighting, let the
unnormalized relative probabilities $p_j(3)$ be the expectation values
of the numbers $N_j$ of times the observation $O_j$ occurs, so
\begin{eqnarray}
p_j(3) = \sum_L \langle\psi|\mathbf{P}_j^L|\psi\rangle
         = \sum_{N=1}^\infty\sum_{L=1}^N|a_N|^2 p_{NLj}.
\label{3measure}
\end{eqnarray}
Then, as always, normalize these to get the normalized observational
probabilities
\begin{eqnarray}
P_j(3) = \frac{p_j(3)}{\sum_k p_k(3)}.
\label{3prob}
\end{eqnarray}

(4) For theory $T_4$, corresponding to volume averaging, let the
unnormalized relative probabilities $p_j(4)$ be the expectation values
of the fraction of regions $N_j/N$ in which the observation $O_j$
occurs (out of all $N$ regions, not just out of the regions with
observations), so
\begin{eqnarray}
p_j(4) = \sum_{N=1}^\infty\frac{1}{N}\sum_{L=1}^N|a_N|^2 p_{NLj},
\label{4measure}
\end{eqnarray}
then giving normalized observational probabilities
\begin{eqnarray}
P_j(4) = \frac{p_j(4)}{\sum_k p_k(4)}.
\label{4prob}
\end{eqnarray}

Theory $T_3$ in its sum over $L$ weights each component state
$|\psi_N\rangle$ by the number of observational regions in which the
observation $O_j$ occurs.  On the other hand, theory $T_4$ has an
average over $L$ for each total number $N$ of regions, so that component
states $|\psi_N\rangle$ with larger $N$ do not tend to dominate the
probabilities for observations just because of the greater number of
observation regions within them.  Theory $T_3$ is analogous to volume
weighting in the cosmological measure, and theory $T_4$ is analogous to
volume averaging \cite{cmwvw}.

(5) For theory $T_5$, which might be said to be observational averaging,
let the unnormalized relative probabilities $p_j(5)$ be the expectation
values of the fraction of observations, $f_j = N_j/N_O$, in which the
observation $O_j$ occurs (out of just the $N_O$ regions with
observations in each eigenstate of the total number $N_O$ of all regions
with one observation), so
\begin{equation}
p_j(5) = \sum_{N=1}^\infty|a_N|^2 \sum_{\mu_1,\mu_2,\ldots,\mu_N}
 |b_{\mu_1\mu_2\ldots\mu_N}|^2 \frac{N_j}{N_O} 
\label{5measure}
\end{equation}
with
\begin{equation}
\frac{N_j}{N_O}=\frac{\langle\mu_1\mu_2\ldots\mu_N|
        \mathbf{P}_j^L|\mu_1\mu_2\ldots\mu_N\rangle}
       {\langle\mu_1\mu_2\ldots\mu_N|
       \mathbf{P}_O^L|\mu_1\mu_2\ldots\mu_N\rangle}.
\label{5ratio}
\end{equation}

This then leads to the normalized observational probabilities
\begin{eqnarray}
P_j(5) = \frac{p_j(5)}{\sum_k p_k(5)}.
\label{5prob}
\end{eqnarray}

Theory $T_5$ would correspond to the procedure of collapsing the full
quantum state $|\psi\rangle$ to one of its bird eigenstates
$|\mu_1\mu_2\ldots\mu_N\rangle$ with probability given by the absolute
square of its amplitude, $|a_N|^2 |b_{\mu_1\mu_2\ldots\mu_N}|^2$, and
then saying in that bird eigenstate the probability of the observation
$O_j$ is the corresponding fraction $f_j = N_j/N_O$ of all observations
that are type $j$.

In \cite{mw} I implicitly assumed that theory $T_3$ is the probability
rule in Everett many-worlds quantum theory and that theory $T_5$ is the
probability rule in collapse versions of quantum theory.  Then I pointed
out that in principle from these two different probability distributions
for observations, one could test between these two versions of quantum
theory.  I still think that $T_3$ would be a more natural rule than
$T_5$ in Everett many-worlds quantum theory, and that $T_5$ is the most
natural rule in collapse versions of quantum theory (assuming that the
collapse is to a bird eigenstate with a definite observation, or none at
all, in each region), but I no longer believe that $T_3$ is the only
possible rule within Everett many-worlds quantum theory.  Although $T_5$
would naturally arise within collapse versions of quantum theory, it is
logically possible it could also arise in many-worlds versions. 
Therefore, whereas I would now say that finding that $T_3$ (or $T_4$)
gives much higher probability for our observation than $T_5$ would tend
to support many-worlds over collapse, it is no longer obvious to me that
finding that $T_5$ gives much higher probabilities for our observation
than $T_3$ or $T_4$ would necessarily support collapse versions of
quantum theory over many-worlds versions without collapse.

Except for theory $T_1$, all of the rules above may be viewed as
modifications of Born's rule, Eq.\ (\ref{Born}), by replacing the
projection operators $\mathbf{P}_j$ with some other {\it observation
operators} $\mathbf{Q}_j(i)$ normalized so that $\sum_j \langle
\mathbf{Q}_j(i) \rangle_i = 1$, giving
\begin{equation}
P_j(i) = \langle \mathbf{Q}_j(i) \rangle_i . 
\label{eventual}
\end{equation}

Of course, one also wants $P_j(i) \geq 0$ for each $i$ and $j$, so one
needs to impose the requirement that the expectation value of each
observation operator $\mathbf{Q}_j(i)$ in each theory $T_i$ be
nonnegative.  The simplest way to do this would be to require that each
observation operator $\mathbf{Q}_j(i)$ be a positive operator.  However,
since a complete theory must both specify the quantum state (here
denoted by $\langle \ldots \rangle_i$, the linear map from operators,
replacing the $\ldots$ in this expression, to complex numbers that are
the quantum expectation values of the operators in the quantum state)
and specify the observation operators $\mathbf{Q}_j(i)$, it is logically
possible that the observation operators $\mathbf{Q}_j(i)$ need not be
positive but just have positive expectation values in the quantum state
$\langle \ldots \rangle_i$ for the same theory $T_i$.

The main point \cite{typder,cmwvw,insuff} of this paper is that in cases
with more than one copy of the observer, such as in a large enough
universe, one cannot simply use the expectation values of projection
operators as the probabilities of observations.  This means that if Eq.\
(\ref{eventual}) is to apply, each theory must assign a set of
observation operators $\mathbf{Q}_j(i)$, corresponding to the set of
possible observations $O_j$, that are {\it not} projection operators,
whose expectation values are used instead as the probabilities of the
observations.  Since these observation operators are not given directly
by the formalism of traditional quantum theory (e.g., as projection
operators by Born's rule), they must be added to that formalism by each
particular complete theory.

In other words, a complete theory $T_i$ cannot be given merely by the
dynamical equations and initial conditions (the quantum state), but it
also requires the set of observation operators $\mathbf{Q}_j(i)$ whose
expectation values are the probabilities of the observations $O_j$ in
the complete set of possible observations by the localized observers
(frogs) within the universe.  (Alternatively, they may be given by some
other rule for the probabilities, if they are not to be expectation
values of operators.)  The probabilities are not given purely by the
quantum state but have their own logical independence in a complete
theory.

For the theories $T_2$--$T_5$, we can write the observation operators
$\mathbf{Q}_j(i)$ in the following forms, with
$\langle\mathbf{Q}\rangle_i = \langle\psi|\mathbf{Q}|\psi\rangle$ when
the quantum state is the pure state $|\psi\rangle$ for all of the
theories $T_i$ under consideration, as we have been taking it to be in
this paper:
\begin{eqnarray}
\mathbf{Q}_j(2) \!\!\!&=&\!\!\!
\frac{\mathbf{P}_j}{\langle\sum_k\mathbf{P}_k\rangle_i},
\nonumber \\ 
\mathbf{Q}_j(3) \!\!\!&=&\!\!\! \frac{\sum_L\mathbf{P}_j^L}
{\langle\sum_k\sum_L\mathbf{P}_k^L\rangle_i},
\nonumber \\  
\mathbf{Q}_j(4) 
\!\!\!&=&\!\!\! \frac{\sum_{N=1}^\infty\frac{1}{N}
            \sum_{L=1}^N\mathbf{P}_N\mathbf{P}_j^L\mathbf{P}_N}
         {\langle\sum_k\sum_{N=1}^\infty\frac{1}{N}
           \sum_{L=1}^N\mathbf{P}_N\mathbf{P}_k^L\mathbf{P}_N\rangle_i},
\nonumber \\  
\mathbf{Q}_j(5) \!\!\!&=&\!\!\!
 \frac{\sum_{N=1}^\infty\sum_{N_O=1}^N\sum_{L=1}^N\frac{1}{N_O}
	    \mathbf{P}_{NN_O}\mathbf{P}_j^L\mathbf{P}_{NN_O}}
 {\langle\sum_k\sum_{N=1}^\infty\sum_{N_O=1}^N\sum_{L=1}^N\frac{1}{N_O}
	    \mathbf{P}_{NN_O}\mathbf{P}_k^L\mathbf{P}_{NN_O}\rangle_i},
	    \nonumber \\
\label{obsop}
\end{eqnarray}
where $\mathbf{P}_N = |\psi_N\rangle\langle\psi_N|$ is the projection
operator onto the component state with $N$ total regions, and
$\mathbf{P}_{NN_O}$ is the projection operator onto the state with $N$
total regions and $N_O$ regions with observations.  The expectation
values of the numerators of these expressions are the relative
probabilities $p_j(i)$, and the expectation values of the full
expressions are the normalized probabilities $P_j(i) \equiv P(O_j|T_i)$
for the observations $O_j$ made by the localized observers (frogs)
within the universe that do not have access to the bird's eye view of
where they are within the universe or of how many observations of the
various types occur within it.

Besides $T_1$, we can have other theories in which the observational
probabilities $P_j(i)$ are not given by the expectation value of any
observation operators $\mathbf{Q}_j(i)$ chosen independent of the
quantum state (up to normalization). For example, the relative
probabilities $p_j(i)$ may be given nonlinearly in terms of quantum
expectation values.

(6) For theory $T_6$, let
\begin{equation}
p_j(6) = \langle\mathbf{P}_j\rangle^c
 = \langle\psi|[\mathbf{I} - \prod_L (\mathbf{I} - \mathbf{P}_j^L)]
    |\psi\rangle^c = p_j(2)^c
\label{6measure}
\end{equation}
and then get the normalized observational probabilities
\begin{eqnarray}
P_j(6) = \frac{p_j(6)}{\sum_k p_k(6)},
\label{6prob}
\end{eqnarray}
where the exponent $c$ is an arbitrary positive constant here and below.

(7) For theory $T_7$, let
\begin{equation}
p_j(7) = \langle\sum_L\mathbf{P}_j^L\rangle^c
       = \langle\psi|\sum_L\mathbf{P}_j^L|\psi\rangle^c = p_j(3)^c,
\label{7measure}
\end{equation}
and then
\begin{eqnarray}
P_j(7) = \frac{p_j(7)}{\sum_k p_k(7)}.
\label{7prob}
\end{eqnarray}

(8) For theory $T_8$, let
\begin{equation}
p_j(8) = \langle\sum_L\sum_{N=1}^\infty\frac{1}{N}
            \sum_{L=1}^N\mathbf{P}_N\mathbf{P}_j^L\mathbf{P}_N\rangle^c
	     = p_j(4)^c,
\label{8measure}
\end{equation}
and then
\begin{eqnarray}
P_j(8) = \frac{p_j(8)}{\sum_k p_k(8)}.
\label{8prob}
\end{eqnarray}

(9) For theory $T_9$, let
\begin{equation}
p_j(9) = \langle\sum_{N=1}^\infty\sum_{N_O=1}^N\sum_{L=1}^N\frac{1}{N_O}
	    \mathbf{P}_{NN_O}\mathbf{P}_j^L\mathbf{P}_{NN_O}\rangle^c
	     = p_j(5)^c,
\label{9measure}
\end{equation}
and then
\begin{eqnarray}
P_j(9) = \frac{p_j(9)}{\sum_k p_k(9)}.
\label{9prob}
\end{eqnarray}

The theories $T_6$--$T_9$ are the nonlinear generalizations of
$T_2$--$T_5$ respectively and reduce to those linear probability rules
for $c=1$.

Let us write what these nine rules would give for the quantum state of
Eq.\ (\ref{example-state}), $|\psi\rangle = \cos{\theta}
|m_0;m_1;m_2\rangle  +\sin{\theta} |n_0;n_1;n_2\rangle$.  Let me also
choose numbers for the parameters of this state to represent very
crudely an equal-amplitude ($\cos{\theta} = \sin{\theta} = 1/\sqrt{2}$)
superposition of the present observable part of the universe and of what
that part may become after a time of $10^{56}$ times its present age,
where the precise value of this time is not important but here is taken
as the time by which it would be probable for our asymptotically de
Sitter region to have decayed if one uses the first number in Eq.\ (89)
of \cite{DGLNSV} as the decay rate per four-volume.

A human brain has a volume of about $10^{101}$ Planck volumes, and the
observable universe today has a volume of about $10^{185}$ Planck
volumes, or about $10^{84}$ times the volume of a human brain, so if we
divide up the present universe into brain-sized regions, we will get of
the order of $10^{84}$ regions, of which of the order of $10^{10}$ are
occupied by human brains.  I shall let $O_1$ represent an ordinary human
observation, and $O_2$ represent an observation by a Boltzmann brain
\cite{DKS,Albrecht,AS,Page05,YY,Page06a,BF,Page06b,Linde06,Page06c,
Vil06,Page06d,Vanchurin,Banks,Carlip,HS,GM,Giddings,typdef,LWc,BS,DP07,
Bou08,BFYa,ADSV,Gott,typder,FL,MA,NYTimes,cmwvw}.  The probability per
human brain volume of a Boltzmann brain is very low, say $10^{-10^{42}}$
\cite{Page06b} for concreteness, so this probability is negligible in
the present observable part of the universe.  Therefore, let us say that
the number of regions of the first component of the quantum state (the
present observable universe) with no observations is $m_0 = 10^{84}$,
the number with an ordinary human observation $O_1$ is $m_1 = 10^{10}$,
and the number with a Boltzmann brain observation $O_2$ is $m_2 = 0$.

The second component of the quantum state, roughly $10^{56}$ Hubble
times to the future, will have a volume (assuming an asymptotically
exponential de Sitter growth at the rate given by the present value of
the cosmological constant) very roughly $10^{10^{56}}$ times that of a
human brain, so I shall take the number of regions without any
observations then to be $n_0 = 10^{10^{56}}$.  Stars and ordinary
observers will have almost entirely died out by then, so I shall take
the number of ordinary human observations then to be $n_1 = 0$. 
Boltzmann brains might occupy a fraction of the universe then that is
very roughly $10^{-10^{42}}$, but by then the universe will have
expanded so enormously large that the total expected number of Boltzmann
brains will be huge, giving the number of Boltzmann brain observations
to be, say, $n_2 = 10^{10^{56}-10^{42}}$.  For simplicity I am ignoring
all other observations but the ordinary human and Boltzmann brain
observations $O_1$ and $O_2$ respectively.  Note that here I have
abandoned the generic case by setting $m_2 = n_1 = 0$.

Now let us see what the nine rules, theories $T_1$--$T_9$, give for the
probabilities $P_1(i)$ and $P_2(i)$ for observations $O_1$ and $O_2$ for
each of the nine value of $i$.

(1) Theory $T_1$, that each observation has equal probability if it has
a nonzero amplitude, gives unnormalized relative probabilities $p_1(1) =
p_2(1) = 1$ and normalized observational probabilities $P_1(1) = P_2(1)
= 1/2$, the same for ordinary human and Boltzmann brain observations. 
If one considered all possible observations rather than just these two,
theory $T_1$ would assign a very low probability for all of them, so the
likelihood that this theory is right would be very low, where
`likelihood' is used with the standard technical meaning of being the
conditional probability $P_j(i) \equiv P(O_j|T_i)$ of our observation
$O_j$, conditional upon the theory $T_i$.

(2) Theory $T_2$, that each observation has a relative (frog)
probability that is the same as the bird probability for that
observation to exist at least somewhere, would give $p_1(2) =
\cos^2{\theta}$ (since only the first component, with amplitude
$\cos{\theta}$, gives rise to the human observation $O_1$ and then with
certainty if the state were that component) and $p_2(2) =
\sin^2{\theta}$ (since only the second component, with amplitude
$\sin{\theta}$, gives rise to the Boltzmann brain observation $O_2$ and
then with certainty if the state were that component).  Since these are
already normalized (because of the special case that $m_2 = n_1 = 0$),
the normalized frog observational probabilities are the same.  For our
choice of equal amplitudes, $\cos{\theta} = \sin{\theta} = 1/\sqrt{2}$,
we get $P_1(2) = P_2(2) = 1/2$, in this case the same as theory $T_1$. 
If one considered all possible observations and labeled our observation
as $O_1$, then because there may be some observations that are unlikely
to occur anywhere, one would have $P_1(2)$ somewhat larger than $P_1(1)$
from theory $T_1$, but still $P_1(2)$ is likely to be so small that
theory $T_2$, as well as theory $T_1$, would be assigned very low
likelihood as a result of our observation.

(3) Theory $T_3$, that the relative frog observation probabilities are
the expected numbers of such observations (the analogue of volume
weighting in the cosmological measure problem \cite{cmwvw}) would give
$p_1(3) = m_1 \cos^2{\theta} + n_1 \sin^2{\theta} = 0.5\times 10^{10}$
and $p_2(3) = m_2 \cos^2{\theta} + n_2 \sin^2{\theta} = 0.5\times
10^{10^{56}-10^{42}}$.  Then the normalized probabilities would be
\begin{eqnarray}
P_1(3) &=& \frac{m_1\cos^2{\theta} + n_1 \sin^2{\theta}}
{(m_1+m_2)\cos^2{\theta} + (n_1+n_2)\sin^2{\theta}} \nonumber \\
&\sim& 10^{-(10^{56}-10^{42}-10)}, \nonumber \\
P_2(3) &=& \frac{m_2\cos^2{\theta} + n_2 \sin^2{\theta}}
{(m_1+m_2)\cos^2{\theta} + (n_1+n_2)\sin^2{\theta}} \nonumber \\
&\approx& 1.
\label{3example}
\end{eqnarray}

Theory $T_3$ has the huge numerical dominance of the expectation value
for the number of Boltzmann brains result in their huge dominance for
the observational probabilities, so that the probability for the human
observation $O_1$ is minuscule.  Our ordinary human observation would
give extremely low likelihood to this theory, so unless one gave it very
nearly all of the prior probability to be correct before considering our
observation, it would be statistically ruled out by our observation at
an enormously high confidence level.  This is a manifestation of the
Boltzmann brain problem
\cite{DKS,Albrecht,AS,Page05,YY,Page06a,BF,Page06b,Linde06,Page06c,
Vil06,Page06d,Vanchurin,Banks,Carlip,HS,GM,Giddings,typdef,LWc,BS,DP07,
Bou08,BFYa,ADSV,Gott,typder,FL,MA,NYTimes,cmwvw}.

(4) Theory $T_4$, that the relative frog observation probabilities are
the expectation values of the fraction of regions in which in
observation occurs (the analogue of volume averaging in the cosmological
measure problem \cite{cmwvw}) would give the relative probabilities
\begin{eqnarray}
p_1(4) &=& \frac{m_1\cos^2{\theta}}{m_0+m_1+m_2}
          +\frac{n_1\sin^2{\theta}}{n_0+n_1+n_2} \nonumber \\
       &\sim& 0.5\times 10^{-74}, \nonumber \\
p_2(4) &=& \frac{m_2\cos^2{\theta}}{m_0+m_1+m_2}
          +\frac{n_2\sin^2{\theta}}{n_0+n_1+n_2} \nonumber \\
       &\sim& 0.5\times 10^{-10^{42}}.
\label{4relprobexample}
\end{eqnarray}
This then gives the normalized probabilities
\begin{eqnarray}
P_1(4) &=& \frac{\frac{m_1\cos^2{\theta}}{m_0+m_1+m_2}
                 +\frac{n_1\sin^2{\theta}}{n_0+n_1+n_2}}
	        {\frac{(m_1+m_2)\cos^2{\theta}}{m_0+m_1+m_2}
                 +\frac{(n_1+n_2)\sin^2{\theta}}{n_0+n_1+n_2}}
       \nonumber \\
       &\approx& 1, \nonumber \\
P_2(4) &=& \frac{\frac{m_2\cos^2{\theta}}{m_0+m_1+m_2}
                 +\frac{n_2\sin^2{\theta}}{n_0+n_1+n_2}}
	        {\frac{(m_1+m_2)\cos^2{\theta}}{m_0+m_1+m_2}
                 +\frac{(n_1+n_2)\sin^2{\theta}}{n_0+n_1+n_2}}
       \nonumber \\
       &\sim& 10^{-(10^{42}-74)}.
\label{4probexample}
\end{eqnarray}

In theory $T_4$ the volume averaging greatly suppresses the Boltzmann
brains, since they are so dilute in the far future component of the
quantum state, much more dilute than human brains are in our present
component of the quantum state.  Therefore, almost all of the normalized
probability is for the ordinary human observation $O_1$.  Of the simple
theories $T_1$--$T_5$ that do not have the arbitrary exponent $c$, it
would seem that the volume-averaged $T_4$ gives the most hope for giving
the highest probability for our observation (the likelihood of the
theory, to be multiplied by the prior probability of the theory to get
the relative posterior probability of the theory in a Bayesian
analysis).

(5) Theory $T_5$, that the relative frog observation probabilities are
the expectation values of the fractions of all observations that are of
the particular type (what might be said to be observational averaging,
or what would most naturally be given by collapse versions of quantum
theory), would give the relative probabilities
\begin{eqnarray}
p_1(5) &=& \frac{m_1\cos^2{\theta}}{m_1+m_2}
          +\frac{n_1\sin^2{\theta}}{n_1+n_2} \nonumber \\
       &=& 0.5, \nonumber \\
p_2(5) &=& \frac{m_2\cos^2{\theta}}{m_1+m_2}
          +\frac{n_2\sin^2{\theta}}{n_1+n_2} \nonumber \\
       &=& 0.5.
\label{5relprobexample}
\end{eqnarray}
This then gives the normalized probabilities
\begin{eqnarray}
P_1(5) = P_2(5) = \frac{1}{2}.
\label{5probexample}
\end{eqnarray} 

This theory $T_5$ is ambivalent whether it makes ordinary human
observations or Boltzmann brain observations more probable.  For the
particular example used here, it would give a likelihood very nearly
half that of theory $T_4$, so both of these theories might seem to be
good candidate theories, with likelihoods not too small.  However, if
one extended the example to an enormous superposition of many different
quantum states, I would suspect that most of probability assigned by
theory $T_5$ would go to the bulk of the quantum components in which it
seems likely that life would be much more rare than it is in our
component, simply because the conditions are much less conducive for
life there.  But in the components in which the conditions are much less
conducive for life, I would suspect that life would be quite different,
and nearly all of the observations would have a significantly different
character than ours do.  Then it would seem that our observations would
have much lower probabilities in theory $T_5$ than in theory $T_4$.

In collapse versions of quantum theory, in which the rule $T_5$ would
arise quite naturally, it would seem most probable that the quantum
state would collapse to one of the presumably many more components in
which life is very rare (on a per-volume basis), and in which most life
that does occur is much very different from ours.  Therefore, it seems
likely that our particular observations would be extremely improbable in
this scenario, much more improbable than they would be in a component of
the quantum state like ours in which many effective coupling constants
and other parameters of the local state seem finely tuned for life and
in which life is presumably not nearly so rare as it would be if the
parameters of our component were not so suitable for life.

Of course, life might be extremely rare, on a per-volume basis or even a
per-planet basis, for all of the components of the quantum state, but
that need not affect the normalized probabilities of our observations,
since one is necessarily selecting for an observation rather than just
randomly selecting a region of space which may or may not have an
observation.  How rare our type of observation is, out of the volume of
space or out of the number of planets, rather than out of all types of
observations that occur with similar probabilities, does not affect the
likelihoods that our observations impute to the theories that predict
these probabilities.  Therefore, it is of no disadvantage to a theory to
predict that life, even within the component of the quantum state that
is most conducive to life, is extremely rare on a per-volume basis.

What does seem likely to make the probability of our observation small
in the actual quantum state of the universe (whatever reasonable, i.e.,
simple, rule is used to deduce the probabilities from the quantum state)
is the fact that the quantum state seems likely to support a large
number of different observations with roughly equal number, spatial
frequency, or frequency among the set of all observations.  However, one
might expect that this number, although no doubt large, is not nearly so
large as the set of all possible observations, so that the probability
of an observation within this dominant subset would be much larger than
the probability of an observation chosen at random, with nearly equal
probabilities, from the set of all possible observations, as theories
$T_1$ and $T_2$ seem likely to give.

In theory $T_4$ one could partially explain the apparent fine tuning as
the selection effect of having the probabilities weighted by the density
of observations per volume (at least if the ultimate theory allowed this
fine tuning to occur in some components of the quantum state, which
itself might be a nontrivial requirement suggesting design).  (The
theory $T_3$ could also partially explain the apparent fine tuning if it
did not have the Boltzmann brain problem.)  Theory $T_5$ does not
incorporate this selection effect, so it would seem that if we want to
explain the apparent fine tuning within our part of the universe (though
not yet explaining an apparent design of the complete theory of physics
that we hope shall predict quantum components with the local fine
tuning), we should reject theories $T_1$, $T_2$, and $T_5$.  If we also
reject theories $T_6$--$T_9$ because of the extra complication of their
exponent $c$ and consider only theories $T_3$ and $T_4$, then it seems
that the Boltzmann brain problem may lead us to prefer theory $T_4$,
which is indeed what I am advocating.  On the other hand, perhaps there
is some other solution of the Boltzmann brain problem that is less
complicated than revising the very simple $T_3$ to the slightly more
complicated $T_4$, in which case one might be able to stick with $T_3$. 
However, so far I have not seen other solutions to the Boltzmann brain
problem that seem simpler than changing the volume-weighted $T_3$ to the
volume-averaged $T_4$.

Although the theories $T_6$--$T_9$, with their arbitrary exponent $c$,
seem uglier than the theories $T_1$--$T_5$ (so that I personally would
assign them lower prior probabilities), they are further proof of the
existence of other possible rules for extracting normalized frog
observational probabilities from the quantum state, so let us continue
the discussion to see what they assign for the two-component quantum
state given by Eq.\ (\ref{example-state}).

(6) Theory $T_6$, the nonlinear analogue of $T_2$, gives the
unnormalized relative probabilities $p_1(6) = p_1(2)^c =
\cos^{2c}{\theta} = 2^{-c}$ and $p_1(6) = p_1(2)^c = \cos^{2c}{\theta} =
2^{-c}$, so for our quantum state with both components having equal
amplitude, $T_6$ gives $P_1(6) = P_2(6) = 1/2$, exactly the same as
theory $T_2$.  However, the probabilities would be different from those
of $T_2$ if we had chosen a state with $\cos^2{\theta} \neq
\sin^2{\theta}$.

(7) Theory $T_7$, the nonlinear analogue of the volume-weighted $T_3$,
gives the relative probabilities $p_1(7) = (m_1 \cos^2{\theta} + n_1
\sin^2{\theta})^c = (0.5\times 10^{10})^c$ and $p_2(7) = (m_2
\cos^2{\theta} + n_2 \sin^2{\theta})^c = (0.5\times
10^{10^{56}-10^{42}})^c$.  Then the normalized probabilities would be
$P_1(7) \sim 10^{-(10^{56}-10^{42}-10)c}$ and $P_2(7) \sim 1$, again
enormously favoring Boltzmann brains over ordinary human brains unless
$c$ were taken to be extraordinarily small, in which case the
probabilities would revert to those of the highly nonpredictive theory
$T_1$ in the limit of $c$ becoming arbitrarily small.

(8) Theory $T_8$, the nonlinear analogue of the volume-averaged $T_4$,
gives the relative probabilities
\begin{eqnarray}
p_1(8) &=& \left(\frac{m_1\cos^2{\theta}}{m_0+m_1+m_2}
          +\frac{n_1\sin^2{\theta}}{n_0+n_1+n_2}\right)^c \nonumber \\
       &\sim& (0.5\times 10^{-74})^c, \nonumber \\
p_2(8) &=& \left(\frac{m_2\cos^2{\theta}}{m_0+m_1+m_2}
          +\frac{n_2\sin^2{\theta}}{n_0+n_1+n_2}\right)^c \nonumber \\
       &\sim& (0.5\times 10^{-10^{42}})^c.
\label{8relprobexample}
\end{eqnarray}
Normalizing these then gives the normalized probabilities
\begin{eqnarray}
P_1(8) &\approx& 1, \nonumber \\
P_2(8) &\sim& 10^{-(10^{42}-74)c}.
\label{8probexample}
\end{eqnarray}

Again, unless $c$ is extraordinarily tiny, this theory $T_8$ very
strongly favors human observations over Boltzmann brain observations in
this toy model.  Presumably $T_8$ would give likelihoods close to what
$T_4$ gives if $c$ is close to 1.  However, other than as a illustration
of the freedom that one has in choosing the rules for extracting the
frog observational probabilities from the quantum state, I do not see
much motivation for complicating the simple linear volume-averaged
theory $T_4$ by going to its nonlinear generalization $T_8$, though it
is conceivable that a further analysis might show that it gives
significantly higher probabilities for our observations, for some
suitable $c$, than the theory $T_4$ that effectively has $c=1$.

(9) Finally, theory $T_9$, the nonlinear analogue of the
observationally-averaged or quantum collapse theory $T_5$, gives the
unnormalized relative probabilities of the frog observations as
\begin{eqnarray}
p_1(9) &=& \left(\frac{m_1\cos^2{\theta}}{m_1+m_2}
          +\frac{n_1\sin^2{\theta}}{n_1+n_2}\right)^c \nonumber \\
       &=& 2^{-c}, \nonumber \\
p_2(9) &=& \left(\frac{m_2\cos^2{\theta}}{m_1+m_2}
          +\frac{n_2\sin^2{\theta}}{n_1+n_2}\right)^c \nonumber \\
       &=& 2^{-c}.
\label{9relprobexample}
\end{eqnarray}
Just as the theory $T_5$ did, for our particular two-component quantum
state $T_9$ gives equal normalized probabilities for the observations,
\begin{eqnarray}
P_1(9) = P_2(9) = \frac{1}{2},
\label{9probexample}
\end{eqnarray} 
which again is not very informative and plausibly gives a very low
likelihood for this theory when one goes to a more realistic model with
a huge number of possible observations.

\section{Conclusions}

The examples show that there is not just one unique rule for getting
observational probabilities from the quantum state.  It remains to be
seen what the correct rule is.  Of the four examples given above with
linear probability rules (relative probabilities given by the first
power of the expectation values of certain operators), that is, theories
$T_2$--$T_5$, I suspect that with a suitable quantum state, theory $T_4$
would have the highest likelihood $P_j(i)$, given our actual
observations, since theory $T_2$ would have the normalized probabilities
nearly evenly distributed over a huge number of possible observations,
theory $T_3$ seems to be plagued by the Boltzmann brain problem
\cite{cmwvw}, and theory $T_5$ would seem to favor components of the
quantum state much more hostile to life than ours and hence probably
having the dominant form of observers quite different from us.  Theories
$T_6$--$T_9$ can be arbitrarily close to $T_2$--$T_5$ respectively if
$c$ is arbitrarily close to 1, but with their arbitrary constant $c$,
they seem more complex than $T_2$--$T_5$ and hence might naturally be
assigned lower prior probabilities.  One might conjecture that the
fairly simple theory $T_4$ can be implemented in quantum cosmology to
fit observations better than other alternatives \cite{cmwvw}.

Thus we see that in a universe with the possibility of multiple copies
of an observer, observational probabilities are not given purely by the
quantum state, but also by a rule to get them from the state.  There is
logical freedom in what this rule is (or in what the observation
operators $\mathbf{Q}_j(i)$ are if the rule is that the probabilities
are the expectation values of these operators).  In cosmology, finding
the correct rule is the measure problem.  Preliminary evidence suggests
that the volume-averaged rule $T_4$ is the best possibility considered
so far.

\section*{Acknowledgments}

I am grateful for discussions with Andreas Albrecht, Tom Banks, Raphael
Bousso, Sean Carroll, Brandon Carter, Alan Guth, Daniel Harlow, James
Hartle, Gary Horowitz, Andrei Linde, Seth Lloyd, Juan Maldacena, Donald
Marolf (who suggested a proof that Born's rule does not work), Mahdiyar
Noorbala, Mark Srednicki, Alex Vilenkin, Alexander Westphal, an
anonymous referee, and others, and especially for a long email debate
with Hartle and Srednicki over typicality that led me to become
convinced that there is logical freedom in the rules for getting
observational probabilities from the quantum state.  This research was
supported in part by the Natural Sciences and Engineering Research
Council of Canada.

\end{document}